\documentclass[11pt]{article}
\usepackage{graphicx}
\def\D{\Delta}
\def\d{\delta}
\def\L{\Lambda}
\def\l{\lambda}
\def\S{\Sigma}
\def\G{\Gamma}
\def\g{\gamma}
\def\e{\epsilon}
\def\s{\sigma}
\def\S{\Sigma}
\def\o{\omega}
\def\O{\Omega}

\def\a{\alpha}
\def\b{\beta}

\def\m{\mu}
\def\n{\nu}
\def\r{\rho}
\def\s{\sigma}
\def\p{\pi}

\def\f{\phi}
\def\F{\Phi}
\def\th{\theta}
\def\t{\tau}
\def\e{\epsilon}

\def\vf{\varphi}

\def\ch{{\cal H}}
\def\cd{{\cal D}}

\def\cl{{\cal L}}

\def\pa{\partial}

\def\det{\textrm{det}}

\def\iu{\textrm{i}}
\def\R{{\bf R}}

\def\bv{{\Big |}}

\newcommand{\be}{\begin{equation}}
\newcommand{\ee}{\end{equation}}
\newcommand{\bea}{\begin{eqnarray}}
\newcommand{\eea}{\end{eqnarray}}

\begin{document}

\begin{center}
\bf{\Large  Finiteness of piecewise flat quantum gravity with matter}
\end{center}

\bigskip
\begin{center}
{\bf\large Aleksandar Mikovi\'c} \footnote{Work supported by national funds from FCT through the project UIDB/00208/2020 
.} \\
\bigskip
Departamento de Inform\'atica e Sistemas de Informac\~ao \\
Universidade Lus\'ofona\\
Av. do Campo Grande, 376, 1749-024 Lisboa, Portugal\\
and\\
Mathematical Physics Group\\
 Instituto Superior T\'ecnico\\
Av. Rovisco Pais, 1749-016 Lisboa, Portugal
\end{center}

\centerline{E-mail: amikovic@ulusofona.pt}

\begin{abstract} 
We review the approach to quantum gravity which is based on the assumption that the short-distance structure of the spacetime is given by a piecewise flat manifold corresponding to a triangulation of a smooth manifold. We then describe the coupling of the Standard Model to this quantum gravity theory and show that the corresponding path integral is finite when the negative power of the product of the edge lengths squared in the path-integral measure is chossen to be grater than 52,5. The implications of this result are discussed, which include a relationship between the effective action and a wavefunction of the universe, the existence of the non-perturbative effective action, the correct value of the cosmological constant and the natural appearence of the Starobinsky inflation.
\end{abstract}

\section{Introduction}

The simplest approach to formulating a quantum theory of gravity is to promote the spacetime metric into a Hermitian operator on a given 4-manifold $M$, which has a simple topolgy of $\S\times \R$, and then apply the canonical or the quantum field theory (QFT) quantization procedure in a straightforward manner, see \cite{Ish}. However, as is well known, the straightforward quantization approach fails. By modifying the straightforward canonical quantization procedure, one can make some advances, as in the case of Loop Quantum Gravity (LQG) \cite{lqg}, but these advances are still not sufficent to construct a satisfactory quantum gravity (QG) theory. In the QFT approach, the problem of the perturbative non-renormalizability of General Relativity (GR) can be avoided by trying to implement a non-perturbative renormalizability based on the asympthotic safety from QFT, which is the Asymptotically Safe Quantum Gravity (ASQG) approach \cite{asqg}. However, there is no a general proof of the existence of the non-perturbative renormalizability in GR QFT, so that ASQG is based on an unproven conjecture.  

In the case of string theory \cite{gsw, pol}, one has a novelty regarding the quantization procedure. Namely, in string theory it is assumed that the short-distance structure of the spacetime is given by a loop super-manifold which is based on a ten-dimensional spacetime manifold. This means that the short-distance (high-energy) structure of the spacetime is not the same as the long-distance one, which is a 4-manifold, and this is the reason why the string theory transition amplitudes are  finite. However, the problem with string theory is that nobody has succeded to recover the Standard Model (SM) in the low-energy limit of the theory. Another problem for string theory is the fact that the observed cosmological constant (CC) is positive, while the string theory is naturally defined for a zero or a negative cosmological constant.

Before string theory and LQG, Regge started the sistematic study of how to construct the GR path integral (PI) by using a piecewise linear (PL) manifold $T(M)$ which is based on a triangulation of a smooth manifold $M$ \cite{regge}. This gave rise to the Regge calculus (RC) approach \cite{RC} and the main difficulty in the RC approach is how to obtain the smooth-manifold limit. In addition, it is not clear how to construct a finite path integral for the PL manifold  $T(M)$, since in the RC approach one considers the Euclidean GR and only a trivial PI measure is used. The RC approach evolved into the spin foam (SF) models \cite{sf} and the causal dynamical triangulations (CDT) approach \cite{cdt}. Both of these approaches have made advances regarding the problem of the path integral finiteness; however, the smooth-manifold limit has been addressed only in the CDT approach. Hence in the SF approach the problem of the smooth-manifold limit remains unsolved, while in the CDT approach, one can only perform numerical simulations, so that there are no exact results.

In this paper we will describe the piecewise flat quantum gravity (PFQG) approach \cite{MVb,M4,M5}, which is based on the Regge path integral. However,  the crucial diference is the structure of the spacetime in the PFQG approach. While in the RC approach the piecewise flat manifold $T(M)$ is considered as an auxilliary tool, which only serves to define the path integral in the smooth-manifold limit, in the PFQG case the PL manifold $T(M)$ is considered as the physical spacetime, which is revaled at short distances. The smooth manifold $M$ is then only an approximation in PFQG, which appears at larger distances. We will also explain how a PFQG theory satisfies the  minimal requirements for a QG theory, i.e. that it is mathematically well-defined and that its classical limit is given by the GR coupled to the SM on a four-dimensional (4d) smooth spacetime, while the quantum corrections are given by the corresponding QFT with a physical cutoff. 

\section{A QG theory definition}

Let us consider a 4d manifold  $M=\S\times [0,t]$, where $\S$ is a 3d manifold. Let $g$ be a metric on $M$ and let $\f$ denote a collection of matter fields on $M$.  A QG theory can be defined  as a map
\be (M, g, \f) \to (\hat M, \hat g, \hat\f) \,, \label{c1}\ee
where $\hat g$ and $\hat\f$ are hermitian operators on some Hilbert space $\ch_\S$ which is associated with the 3-manifold $\S$. $\hat M$ is the quantum spacetime, which could be $M$, or some larger space, as is the case in string theory. Note that in canonical LQG and in ASQG it is assumed that $\hat M = M$,  while in the case of SF models and CDT one starts with $\hat M = T(M)$, but one then has to perform the smooth-manifold limit $T(M) \to M$. In the PFQG case $\hat M = T(M)$.

Beside the map (\ref{c1}), one should also be able to construct an evolution operator $\hat U(t)$ such that
\be |\Psi_\S (t) \rangle = \hat U(t) |\Psi_\S (0) \rangle \,,\label{c2}\ee
where $\Psi_\S \in \ch_\S$.

A satisfactory QG theory should also have the semi-classical states $|\Psi_{\S, p_0, q_0}(t)\rangle$ such that
\be \langle \Psi_{\S, p_0,q_0}(t)|\,\hat q \,|\Psi_{\S, p_0,q_0}(t) \rangle = q_0(t)\left[ 1 + O\left({\hbar\over S_0(t_0)}\right)\right] \,, \ee
and
\be\langle \Psi_{\S, p_0,q_0}(t)|\,\hat p \,|\Psi_{\S, p_0,q_0}(t) \rangle = p_0(t) \left[1+ O\left({\hbar\over S_0(t_0)}\right)\right] \,, \ee
where 
\be S_0(t)= \int_0^t d\t \left(p_0\dot q_0 - H(p_0,q_0,\t)\right)\,, \ee 
is the classical action of our QG theory. Here $(p,q)$ denote the independent canonical coordinates and momenta of the classical action,  while $(p_0(t), q_0(t))$ denotes a solution of the classical equations of motion (EOM). The time $t_0$ is a characteristic timescale of the problem considered.

Now we can describe more precisely the problems of the well-known candidate QG theories. In the case of the standard QFT quantization of GR coupled to SM, the operator $\hat U(t)$ is not well-defined because of perturbative non-renormalizability of GR.  In the case of ASQG approach there is no proof that $\hat U(t)$ exists. In the CDT case, there are only approximative expressions for $\hat U(t)$ and it is not clear what is the exact $\hat U(t)$. The same applies to SF models and canonical LQG. Furthermore, the semiclassical states are not known in LQG, so that it is not clear how to recover the SM QFT.

In the case of string theory, $\hat U(t)$ is well defined perturbatively, and the semiclassical states can be constructed. However, the main problem is how to obtain the correct classical limit, which is GR coupled to the SM.

In the PFQG case, $\hat M = T(M)$ and for a compact $M$ and a finite triangulation the number of degrees od freedom (DOF) is finite, because the DOF are the edge lengths and the matter fields values at the vertices of $T(M)$. When $M$ is a non-compact manifold, one takes the non-zero edge lengths and the field values only at a $T(B\times I)$, where $B$ is a 3-ball and $B\times I \subset M$, see the next section. Consequently $\hat U(t)$ can be defined exactly, since the corresponding path integral is given by a finite-dimensional Riemann integral, which can be made convergent by an appropriate choice of the integration measure \cite{M1,M4}. 

The correct classical and the correct semi-classical limit can be obtained in the PFQG theory when the number of edges is large and the edge lengths are sufficiently small, with an appropriate choice of the PI measure \cite{M,M1}. In this case the PL manifold $T(M)$ is very well approximated by the smooth manifold $M$ and one can use GR QFT coupled to SM QFT with a cutoff $\hbar/L$, where $L$ is the average edge length in $T(M)$. This is analogous to the fluid dynamics approximation when describing the motion of a liquid or a gas. Since a liquid or a gas consists of many molecules, instead of using the Newton equations for a system of a large number of particles, we can introduce the average molecular velocity for a microscopic volume, such that there are many molecules in each microscopic volume. If we are interested in describing the fluid motion at a scale much larger than the microscopic volume size, then we can introduce a smoothly varying fluid velocity field, which is then described by the Navier-Stokes partial differential equations.
 
\section{GR path integral in PFQG}

We will now describe the definition of the PFQG path integral \cite{M, MVb}. Let $M$ be a smooth 4d manifold and let $T(M)$ be a PL manifold corresponding to a regular triangulation of $M$, i.e. the dual one simplex is a connected 5-valent graph. Let us consider
\be  M = M_1 \sqcup \left(\S\times I \right)\sqcup M_2 \,, \label{eat}\ee
where $\S$ is a 3-manifold, $I$ an interval from $\R$ and $M_1$ and $M_2$ 4-manifolds such that $\pa M_1 = \pa M_2 = \S$, see Fig. 1. 
\begin{figure}[htpb] 
\centering
\includegraphics[width=0.6\textwidth]{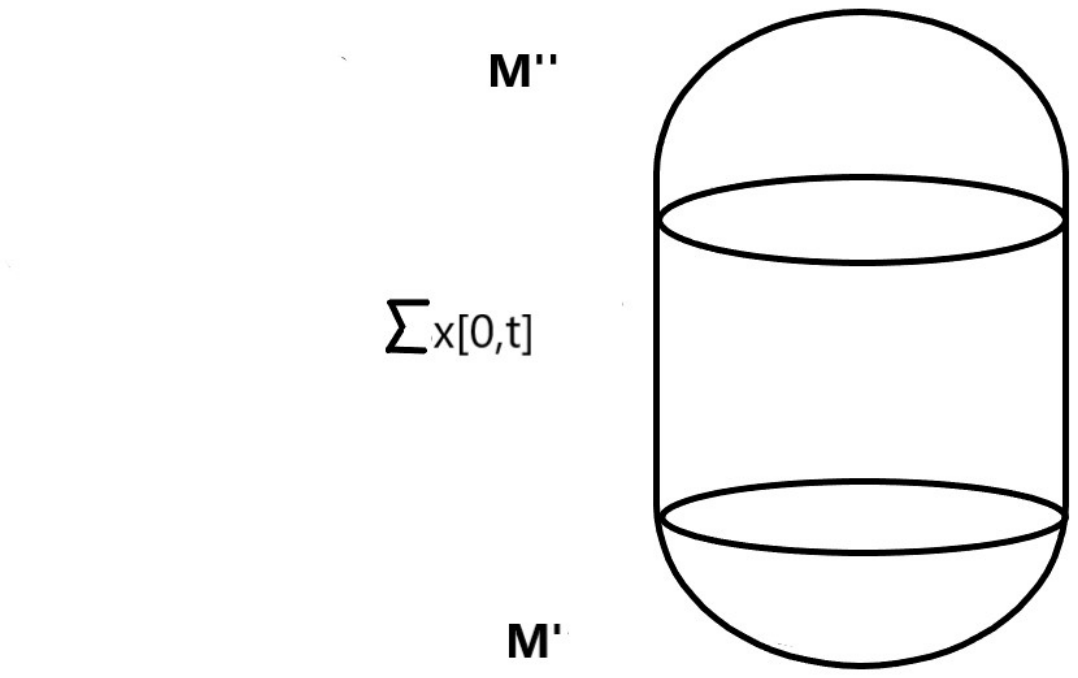}
\caption{Topology of a PFQG closed spacetime manifold}
\end{figure} 

Although we can take $M$ to be of an arbitrary topology, the restriction (\ref{eat}) allows us to make a connection with the canonical quantization as well as to associate a wavefunction of the universe (WFU) with an effective action, see \cite{M5}.

When $\S$ is a non-compact manifold, we maintain a finite number of DOF by allowing non-zero $L_\e$ only for a triangulation of a compact subset of $M$, given by
\be B_4 \sqcup \left(B_3 \times I\right) \sqcup B_4 \,,\ee 
where $B_3$ is a 3-ball in $\S$, while the to two 4-balls in $M_1$ and $M_2$ are glued at the ends of the interval $I$, see Fig. 2.
\begin{figure}[htpb] 
\centering
\includegraphics[width=0.6\textwidth]{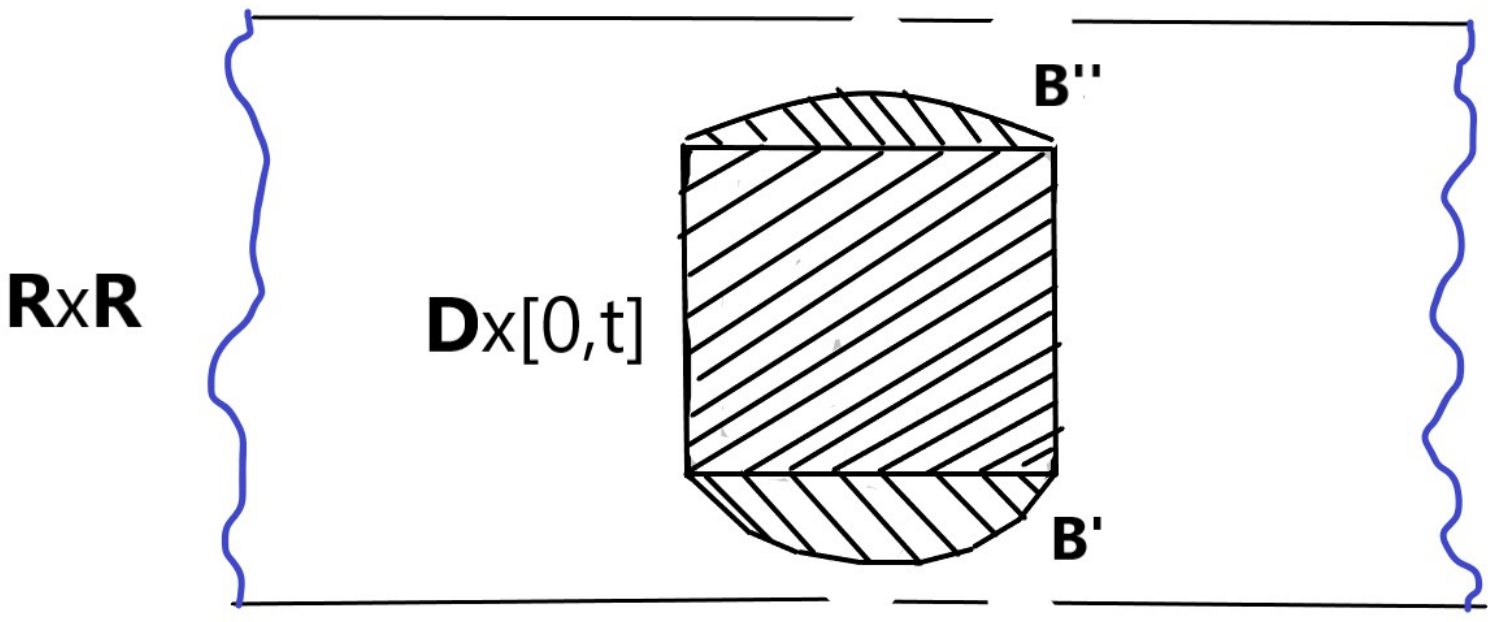}
\caption{Topology of a PFQG non-compact spatial manifold}
\end{figure} 

Let $\{L_\e |\, \e\in T_1(M)\}$ be a set of the edge lengths  such that  $L_\e^2 \in \R$, i.e.
$L_\e \in \R_+$ (a spacelike edge) or $L_\e \in \iu\R_+$ (a timelike edge).

A metric can be introduced on $T(M)$,  which is flat in each 4-simplex $\s$ of $T(M)$, and it is given by 
\be  G_{\m\n}(\s) = L_{0\m}^2 + L_{0\n}^2 - L_{\m\n}^2 \,,\ee
where the five vertices of $\s$ are labeled as $0,1,2,3,4$ and $\m,\n = 1,2,3,4$.  This metric is known as the Cayley-Menger (CM) metric. 

The CM metric is not dimensionless and hence it is not diffeomorphic to 
\be g_{\m\n}(\s) = diag(- 1, 1, 1, 1) \,.\ee 
This can be corrected by using a dimensionless PL metric 
\be g_{\m\n}(\s) = {G_{\m\n}(\s) \over |L_{0\m}||L_{0\n}|} \,,\ee
see \cite{MVb}.

The Einstein-Hilbert (EH) action on $M$ is given by
\be S_{EH} = \int_M \sqrt{|\det g|}\, R(g) \, d^4 x  \,,\ee
where $R(g)$ is the scalar curvature associated to a metric $g$ on $M$. On $T(M)$ the EH action becomes the Regge action 
\be S_R (L) = \sum_{\D\in T(M)} A_\D (L)\, \d_\D (L) \,,\label{era}\ee
when the edge lengths correspond to a Eucledean PL geometry. $A_\D$ is the area of a triangle $\D$, while  the deficit angle $\d_\D$ is given by
\be \d_\D = 2\pi - \sum_{\s \supset \D} \th_\D^{(\s)} \,,\ee
where a dihedral angle $\th_\D^{(\s)}$ is defined as the angle between the 4-vector normals associated to the two tetrahedrons that share the triangle $\D$. 

In the case of a Lorentzian PL geometry, a dihedral angle can take complex values, so that it is necessary to modify the Regge action formula (\ref{era}) such that the Regge action takes only the real values. The appearence of the complex dihedral angles can be seen from the formula
\be \sin \th_\D^{(\s)} = \frac{4}{3} {v_\D v_\s \over v_\t v_{\t'}} \,, \ee
where $v_s = V_s \ge 0$, if the CM determinant is positive, while $v_s = \iu V_s$  if the CM determinant is negative. Consequently, $\sin \th_\D^{(\s)} \in \R$ or $\sin \th_\D^{(\s)} \in \iu{\bf R}$. This implies that the Regge action (\ref{era}) will give a complex number when the spacelike triangles are present. 

One can modify the Regge action as
\be  S_R(L) = Re\left((-\iu)\sum_{\D\in ST} A_{\D} \,\d_{\D} \right) + \sum_{\D\in TT} A_{\D} \,\d_{\D}  \,,\label{lra}\ee
where $ST$ denotes the set of spacelike triangles, while $TT$ denotes the set of timelike triangles \cite{MVb}. Consequently the action (\ref{lra}) is always real and it corresponds to the Einstein-Hilbert action on $T(M)$.

We can now define the GR path integral as
\be  Z(T(M)) = \int_D \prod_{\e=1}^N dL_{\e} \, \m(L) \, e^{i S_R(L)/l_P^2} \,, \ee
where $dL_\e = d|L_\e|$ and $\m(L)$ is a mesure that ensures the finiteness and gives the effective action with a correct semiclassical expansion, see \cite{M,M1}. The integration region $D$ is a subset of ${\bf R}_+^N$, consistent with a choice of spacelike and timelike edges.

The integral $Z(T(M))$ is convergent for the measure
\be \m(L) = e^{-V_4 (L)/L_0^4}\prod_{\e=1}^N \left(1+ {|L_\e|^2 \over l_0^2} \right)^{-p} \,,\label{pim}\ee
where $p > 1/2$ and $V_4$ is the 4-volume of $T(M)$, see \cite{M1}. The bound $p>1/2$ can be easily derived from the requirement of the absolute convergence
\be |Z| \le \int_D \prod_{\e=1}^N dL_{\e} \, \m(L) <  \prod_{\e=1}^N \int_0^\infty dL_{\e} \, \left(1+ {|L_\e|^2 \over l_0^2} \right)^{-p} \,. \label{ac}\ee

Note that (\ref{ac}) implies that the convergence can be also obtained without the $e^{-V_4/L_0^4}$ factor in the measure. However, the exponential factor is necessary in order to obtain the correct classical limit of the effective action. Namely, when $|L_\e| \to\infty$, we need
\be {\pa^2\log\m(L)\over \pa L_\e^2} < 0 \,,\label{scc}\ee
for all $\e$ in $T(M)$, see \cite{M, M1, MVb}. This condition is satisfied by the measure (\ref{pim}), while the measure (\ref{pim}) without the exponential factor does not satisfy the condition (\ref{scc}).

\section{PFQG with the SM matter}

We will now investigate the convergence of the PFQG path integral when the SM matter is coupled to gravity. In this case the SM action on $M$ is given by
\be S_m = S_H + S_{YM}  + S_f  + S_{Y}  = \int_M d^4 x \,\sqrt{g}\left(\cl_H + \cl_{YM} + \cl_f + \cl_Y \right) \,,\label{sma}\ee
where 
\be \cl_H = \frac{1}{2} D^\m \f^\dag  D_\m \f - \l_0^2 (\f^\dag \f - \f_0^2)^2 \,, \ee
is the Higgs field Lagrangian, with $D_\m \f = (\pa_\m  + \iu \,(g_0 A)_\m )\f$ and
\be g_{0} A =   g_{01} A_1 +  g_{02} A_2 + g_{03} A_3 \in u(1)\oplus su(2)\oplus su(3) \,. \ee

The Yang-Mills (YM) Lagrangian is given by
\be \cl_{YM} = -\frac{1}{4}\, Tr\left(F^{\m\n} F_{\m\n} \right) \,,\ee
where the trace is over the SM Lie algebra, while the fermionic matter Lagrangian is given by
\be \cl_f = \sum_{k=1}^{48} \e^{abcd} e_b \wedge e_c \wedge e_d \, \bar\psi_k \left(\iu \g_a ( \textrm{d} + \iu \o +  \iu g_0 A) \right)\psi_k \,, \ee
where $\o$ is the spin connection on $M$.

The Yukawa couplings are described by the Lagrangian 
\be \cl_Y = \sum_{k,l} Y_{kl}\,\langle \bar\psi_k \psi_l \f \rangle\,,\quad \,,\ee
where $\langle\bar\psi_k \psi_l \f\rangle$ is the scalar invariant in the tensor product of the corresponding SM group representations.

On $T(M)$, the smooth-manifold actions in (\ref{sma}) become the corresponding PL actions. We will take as matter DOF the values of the matter fields at the vertices of the triangulation, since it simplifies the analysis of the path-integral convergence. For the Higgs action we obtain
\be \tilde S_H = \sum_{\s\in T_4} V_\s (L) \, s_{HK} + \sum_{\p\in T_0} V_\p (L) \, s_{HP} \,,\ee
where $T_k$ is the set of $k$-symplexes in $T(M)$ and $V_\p$ is the volume of the dual cell to a vertex $\p$, while
\be s_{HK} =g^{\m\n}_\s  \left(\frac{\f(\p_\m) - \f (\p_0)}{|L_{0\m}|} + \iu g_0 A_\m (\p_0) \f_{\p_0} \right)^\dag \left(\frac{\f(\p_\n) - \f (\p_0)}{|L_{0\n}|} + \iu g_0 A_\n (\p_0) \f_{\p_0} \right)\ee
and
\be s_{HP} =    \l_0^2  \left(\f^\dag (\p)\f (\p) - \f_0^2 \right)^2 \,.\ee

The fermion action on $T(M)$ is then given by
\be \tilde S_f = \sum_{\e\in T_1} V_\e ( L) \, s_{f} +\sum_{\p\in T_0} V_\p ( L) \, \,s_{YMf} \,,\ee
where
\be s_{f} = \sum_{k=1}^{48} \e^{abcd} \, B_{abc}(p) \,\bar\psi_k (\p)\,\iu \g_d\,(| L_\e| \,\iu \o_\e (L)\,\psi_k (\p') + \psi_k (\p') - \psi_k (\p)) \,,\ee
\be  s_{YMf} = \sum_{k=1}^{48}  \bar\psi_k (\p) \, g_0 \g^\m (\p) A_{\m}(\p)\, \psi_k (\p) \,.\ee
$V_\e$ is the volume of the dual cell to an edge $\e$,
\be \g^\m (\p) = e^\m_a (\p) \g^a \,, \quad e^\m_a (\p) ={1\over n_4(\p)} \sum_{\s :\, \p\in \s} e^\m_a (\s) \,,\ee
where $\g^a$ are the Gamma matrices and $n_4(\p)$ is the number of 4-simplexes that share the vertex $\p$.

Note that the tetrads on $T(M)$ are naturally associated to the 4-simplices $\s$, since the PL metric is defined in each 4-symplex. The spin connection $\o$ is then defined with respect to the dual edges, so that we have a set of $\o_l$ values, where $l\in T_1^*(M)$. We can then define $\o_\e$ on $T(M)$ as
\be \o_\e = \frac{1}{n_3(\e)}\sum_{\t :\, \e\in\t} \o_{l(\t) }\,,\ee
where $n_3(\e)$ is the number of tetrahedrons that share an edge $\e$.

The Yukawa action on $T(M)$ is then given by
$$\tilde S_Y = \sum_\p V_\p (L)\,  s_Y \,, $$
where
$$s_Y = \sum_{k,l} Y_{kl} \langle\bar\psi_k (\p) \psi_l (\p)\f (\p) \rangle\,.$$

In order to take into account that only $2|G|$ components out of $4|G|$ components per a spacetime point of a YM field are independent, one should also add to $S_m$ the YM ghost action 
\be S_{gh} = \int_M \sqrt{g}\, Tr \left(\pa^\m \bar c \,D_\m c \right) d^4x \,.\ee 

Therefore the PFQG path integral for gravity plus matter can be written as
\be Z = \int_D d^N L \,\mu(L)\, e^{i S_R(L)/l_P^2}\, Z_m(L) \,,\label{gmpi}\ee
where $\m(L)$ is given by (\ref{pim}) and
\be Z_m (L) = \int_{D_m}\prod_{a=1}^{c_b} d^n \f_a  \prod_{\a =1}^{c_f} d^n \bar\psi_\a \, d^n \psi_\a \, e^{i \tilde S_m(\F,\Psi, L)/\hbar} \,.\label{mpi}\ee
Here $\F$ denotes the collection of the SM bosonic field components $\f_a$,  $\Psi$ denotes the collection of the SM fermionic fields $\psi_\a$ (including the YM ghosts) and $n$ is the number of vertices in $T(M)$. One can see that $c_b = 4(1+3+8) + 4 = 52$, while $c_f = 96 + 24 = 120$.

The integration region $D_m = \R^{c_b}$, since the fermionic fields are considered as the generating elements of a Grassman algebra of dimension $2^{2c_f}$, so that the integration is equivalent to a diferentiation\footnote{More precisely, a function of the Grassman coordinates $\th_1,\cdots, \th_n$ can be considered as a vector in a vector space whose basis is given by $2^n$ vectors $\{1, \th_k, \th_k \th_l, \cdots, \th_1\th_2\cdots\th_n \}$, so that
$$ f(\th) = f_0 + f_{k}\, \th_k  + f_{kl} \,\th_k \th_l  + \cdots + f_{12...n}\, \th_1 \cdots \th_{n} \,.$$
Hence one can define $\int d^n\th\, f(\th)$ to be the number $f_{12...n}$, which is the same as diferentiating $f(\th)$ with $\pa_{\th_n} \cdots\, \pa_{\th_1}$.}.

Since the convergence of $Z_m$ is not guaranteed, we pass to a Eucledean geometry defined by the edge lengths
\be \tilde L_\e = |L_\e| \,, \ee
so that all the Eucledean edge lengths are positive real numbers. This is equvalent to a Wick rotation where $\tilde L_\e = L_\e$ if $\e$ is a spacelike edge and $\tilde L_\e = (-\iu)L_\e$, if $\e$ is a timelike edge.

After the Wick rotation the integral (\ref{mpi}) becomes
\be \tilde Z_m ( \tilde L) = \int_{D_m}\prod_a d^n  \f_a \prod_\a d^n\bar\psi_\a\,d^n\psi_\a\, e^{-\tilde S_m(\F, \Psi, \tilde L)/\hbar} \,,\ee
where $\tilde S_m$ is the Euclidian matter action. The integration of the fermions gives a product of the corresponding determinants\footnote{This is due to the relation $$ \int d^n\th\int d^n\bar\th \exp(\bar\th M \th) = \det M \,. $$}, which are polynomial functions of the bosonic fields components. Hence
\be \tilde Z_m ( \tilde L) = \int_{D_m}\prod_a d^n  \f_a\, P(\F, \tilde L)\, e^{-\tilde S_{bm}(\F, \tilde L)/\hbar} \,,\ee
where $P(\F,\tilde L)$ is a polynomial function of $\F$ and $\tilde S_{bm}$ is the Eclidean bosonic action.

Since $\tilde S_{bm} (\F, \tilde L)$ is a polynomial function of $\F$ of degree four, such that $\tilde S_{bm}$ is positive for large $\f_a$,
then the integral $\tilde Z_m$ will be convergent\footnote{This is due to the fact that $\int_{-\infty}^\infty x^n e^{-P_4 (x)} \,dx$ is convergent for any  $n \ge 0$ and any fourth-order polinomial $P_4(x)$ whose coefficient of the $x^4$ term is positive.}. Hence we will define
\be Z_m(L) = \tilde Z_m (\tilde L)\bv_{\tilde L = w(L)} \,,\ee
where $w$ is the Wick rotation.

The SM action on $T(M)$ can be written as
\be S_m = S_1 + S_2 +  S'_2 + S_3 \,, \ee
where
\bea S_1 &=& r^3\langle  \bar\psi\psi \rangle + r^4\langle \bar\psi\psi A \rangle + r^4\langle \bar\psi\psi \f \rangle \nonumber\\
S_2 &=&  r^4 \langle ( Ar^{-1} + g_0 A^2 )^2\rangle   \nonumber\\
S'_2 &=& r^3 \langle \bar c ( r^{-1} + g_0 A c )\rangle \nonumber\\
S_3 &=& r^4 \langle ( \f r^{-1} + g_0 A\f )^2\rangle + \l_0^2 r^4 \langle (\f^2 - \f_0^2)^2\rangle  \,.\label{scd}\eea
A bracket $\langle XY\cdots \rangle$ in (\ref{scd}) represents a sum over the components and the simplex values of the fields $X$, $Y$, ... , so that
\be \langle X Y \cdots \rangle = \sum_{\a,\b,...} c^{\a\b ...}(\th) X_\a Y_\b \cdots \,, \ee
where $(r, \th)$ are the spherical coordinates for a vector $\tilde L = (\tilde L_1, \cdots, \tilde L_N)$.

After integrating the fermions and the ghosts, it can be shown that
\be |Z_m (L)| < r^{c'n} F_n(\th) \,,\ee
where
\be c'=3c_f - c_b^* = 3c_f - 2|G| - 4 = 260\,,\ee 
and
\be F_n(\th) = \int \cd\chi \, \cd\xi\, e^{-s(\th,\xi,\chi)} \D_{ferm} (\xi,\chi) \D_{ghost}(\xi) \,,\ee
see \cite{M4}. The new variables are given by $\xi = rA$ and $\chi = r\f$, while $s(\th,\xi,\chi)$ is the YM action plus the kinetic part of the Higgs action. $\D_{ferm}$ is the fermionic determinat and $\D_{ghost}$ is the ghost determinant. 

Consequently
\be |Z| < \int_D d^N L \,\m(L) |Z_m(L)| < \int d^N L \, \m(L) r^{c'n} F_n(\th) \,,\ee
so that
\be |Z| < \int_0^\infty r^{N-1 +c'n}dr \int_\O J_N(\th) \m(r,\th) F_n(\th)d^{N-1}\th \,,\ee
where $d^N L = r^{N-1} dr J_N(\th)d^{N-1}\th$. By using the asymptotic properties of $\m(r,\th)$ for small $r$ and for large $r$, we obtain
\be |Z| < C_1 \int_0^R r^{c'n + N-1} dr + C_2 \int_R^\infty r^{c'n + N-1 -2pN} dr \,.\ee

Hence we can guarantee the absolute convergence of the PFQG path integral (\ref{gmpi})  if 
\be c'n + N(1-2p) < 0 \,,\ee
so that
\be {c'\over 2p-1} < {N\over n} \,.\ee

For a regular triangulation we have
\be {N\over n} \ge {N_1^*\over N_0^*} \ge \frac{5}{2} \,,\ee
where $N_1^*$ is the number od dual edges and $N_0^*$ is the number of dual vertices, so that if $c'/(2p-1) < 5/2$, then the absolute convergence bound will be satisfied, which gives
\be p > \textrm{52,5} \,.\label{prm}\ee

Hence for the values of the parameter $p$ that satisfy (\ref{prm}) we know that the PFQG path integral (\ref{gmpi}) is convergent.

\section{The effective action}

The concept of the effective action is fundamental for the PFQG theory, since the effective action defines the classical limit and the semi-classical (SC) expansion. The correctness of the SC expansion gives a restriction on the PI measure (\ref{scc}), which combined with the finiteness of the GR path integral leads to the PI measure (\ref{pim}). Then the finiteness of the path integral for the GR plus SM gives a bound (\ref{prm}). 

The PFQG effective action can be defined by using the QFT definition. Namely, a QFT effective action $\G[g,\f]$  can be determined from the EA equation
\be e^{i\G[g,\f]/\hbar} = \int\cd h \cd\vf \exp\left(\frac{\iu}{\hbar}S[g+h, \f + \vf] - \frac{\iu}{\hbar}\int_M \left(\frac{\d\G}{\d g(x)}\,h(x)+ \frac{\d\G}{\d\f(x)}\,\vf(x) \right)\sqrt{g}\, d^4 x \right) \,,\label{seae} \ee
where $S[g,\f]$ is the classical action for GR plus the SM on a smooth spacetime manifold $M$. From this equation one can easilly find the perturbative expansion of the effective action in powers of $\hbar$. However, in order to find a non-perturbative solution, it is essential that the path integral
\be Z = \int \cd g \int \cd \f \exp\left(\frac{\iu}{\hbar} S[g,\f]\right) \,,\ee
is finite, which is realized in the PFQG case.

Since an effective ation should describe a time evolution on $\S\times I$ manifold, this is the reason why we have restricted the topology of $M$ to that given by (\ref{eat}). We also take that $M_1 = M_2$, since this is the way to connect an effective action with a wavefunction of the Universe (WFU), see \cite{M5}. Namely, in this case the WFU is given by the Hartle-Hawking (HH) wavefunction for the manifold $M_1$, see fig. 3, while the interval $I= [-t, t]$ will give the EA trajectories that correspond to the expectation value trajectories in the HH state on the interval $[0,t]$, see fig 4.
\begin{figure}[htpb] 
\centering
\includegraphics[width=0.6\textwidth]{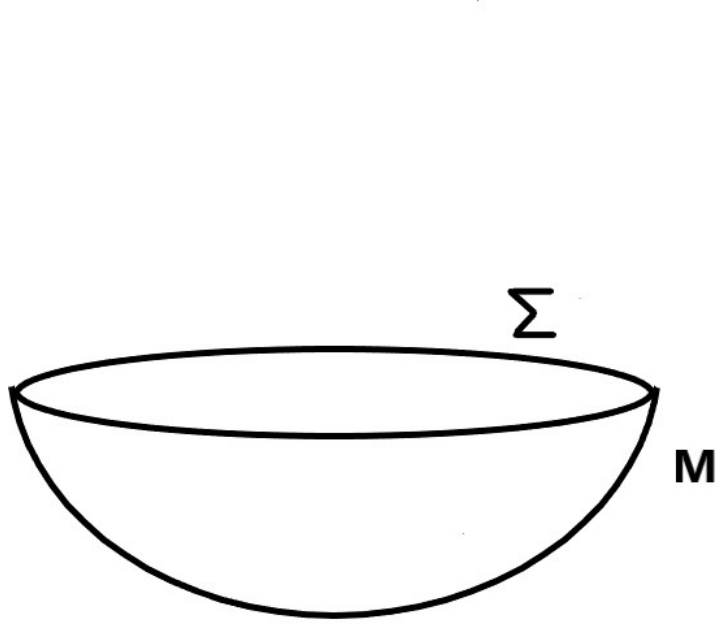}
\caption{Topology of the Hartle-Hawking manifold}
\end{figure} 

\begin{figure}[htpb] 
\centering
\includegraphics[width=0.6\textwidth]{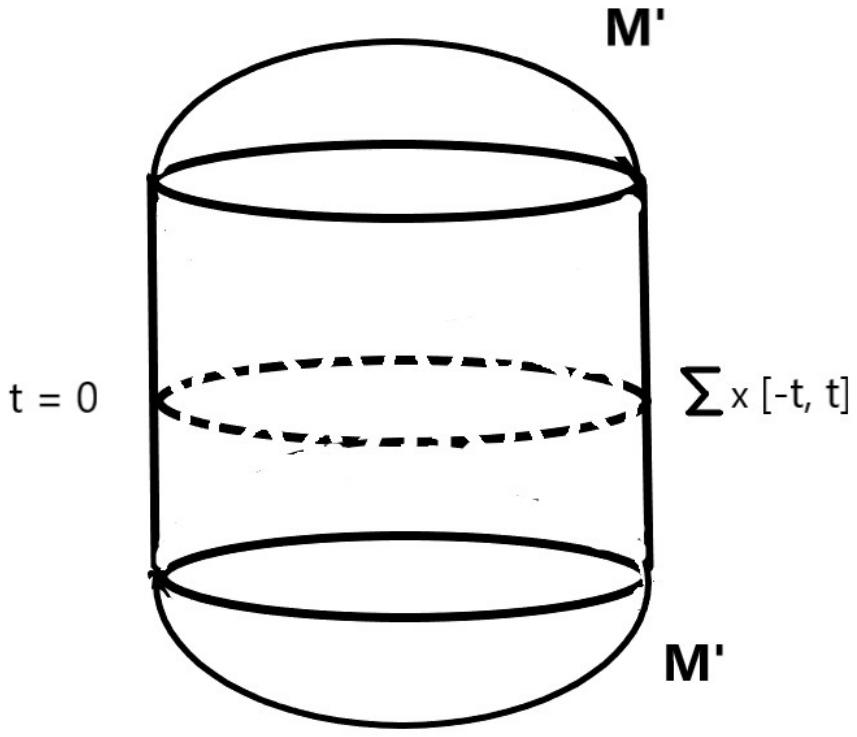}
\caption{Topology of the effective action manifold}
\end{figure} 

On $T(M)$ the EA equation (\ref{seae}) becomes
\be  e^{i\Gamma(L,\F)/\hbar} = \int_{D(L)} d^N  l \int_{D_m}  d^{cn} \vf \, \mu (L +l)\, e^{iS (L+l, \F + \vf)/\hbar - i\sum_{\epsilon} \G'_\epsilon (L,\F)l_\epsilon /\hbar -  i\sum_{\p} {\Gamma}'_\p (L,\F) \vf_\p /\hbar } \,,\label{gmea}\ee
where $c$ is the number of components of the matter fields ($c = c_f + c_{gh} + c_b = 96 + 24 + 52 = 172$ for the SM) and 
\be S(L,\F)= \frac{1}{G_N}S_R(L) + S_m(L,\F) \,.\ee

For the simplicity sake, the vector $\F$ now includes both the bosonic and the fermionic field components, so that
\be d^{cn}\vf = \prod_a d^n \vf_a \prod_\a d^n \psi_\a d^n \bar\psi_\a \,.\ee 

The EA equation will be only defined if the gravity plus matter path integral is finite, which is the case for $p >$ 52,5. This is the consequence of
\be |\tilde Z_m(\tilde L, J)| \le \tilde Z_m(\tilde L) \,,\ee
where
\be \tilde Z_m(\tilde L, J) = \int_{D_m} d^{cn}  \f \, e^{[-\tilde S_m(\F,  \tilde L) + i J\F ] /\hbar} \,,\ee
is the matter path integral with the sources $J$ (the analog of the QFT generating functional).

Note that the EA equations (\ref{seae}) and (\ref{gmea}) will generically give a complex $\G$. In the QFT case without gravity, one can solve this problem via the Wick rotation, since the matter path integral is finite in the case of a Euclidean spacetime metric. However, this approach does not work when the metric DOF are included in the path integral, since the Euclidean GR path integral is not necessarily convergent for the measure (\ref{pim}). This problem can be solved in the following way: if $\G$ is a complex solution of the EA equation, then the real effective action can be defined through a substitution
\be \G \to Re\,\G + Im\,\G \,,\ee
see \cite{M, MVb}.

The effective action is also important for the definition of the smooth-manifold approximation. Let $N\to\infty$ and $|L_\e| =O(1/N)$ in $T(\S\times I)$, such that
\be g_{\m\n}(\s) \approx g_{\m\n}(x) \,, \quad \F_\a (v) \approx \f_\a (x)  \quad  \textrm{for}\,\, x\in\s  \,\, \textrm{and}\,\, v = \textrm{dual vertex}\in\s \,,\ee
where $g_{\m\n}(x)$ is a smooth metric on $\S\times I$ and $\f_\a (x)$ is a smooth matter field on $\S\times I$. Then
\be \G(L,\F) \approx \G_K [g_{\m\n}(x), \f(x)]  \,,\ee
where $\G_K$ is the QFT effective action for GR plus the SM with the cutoff $K = 2\pi/\bar L$ and $\bar L$ is the average edge length in $T(\S\times I)$.

This is a consequence of a theorem that a  Fourier series expansion of a PL function on an interval $N\bar L$ can be approximated by a Fourirer integral with a cutoff $K=2\pi/\bar L$ for $N$ large \cite{MVb}.

The relationship between the PL and the smooth-manifold effective action can be understood from the perturbative expansions in $\hbar$
\be {\G(L,\F) \over\hbar} = {S_{R} (L) + G_N\, S_m (L,\F)\over l_P^2 } +  \G_1 (L,\F) +  l_P^2 \,\G_2 (L,\F) +  l_P^4 \,\G_3 (L,\F) +\cdots \,,\label{lpe} \ee
and
\be \G_K (g,\f)= \frac{1}{G_N}S_{EH} (g) + S_m (g,\f) +  \hbar\G_K^{(1)}(g,\f) + \hbar^2 \,\G_K^{(2)}(g,\f) +  \hbar^3 \,\G_K^{(3)}(g,\f) + \cdots \,,\label{hbe} \ee
where $\G_n$ is determined by the EA equation (\ref{gmea}), while $\G_K^{(n)}$ is the $n$-loop QFT effective action for GR coupled to matter. Note that $l_P^2 = G_N\hbar$ and the expansion (\ref{lpe}) is valid for $|L_\e| \gg l_P$ and small $\f$,  so that 
\be (G_N)^{n-1} \G_n (L,\F) \approx \G_K^{(n)} [g(x), \f(x)]\,,\quad n = 0,1,2, ... \,,\ee 
for $N$ large and $|L_\e| =O(1/N)$, see \cite{MV, M1}. 

Note that $|L_\e| \gg l_P$ still allows for $|L_\e|$ to be microscopically small, so that the smooth-manifold approximation is still valid. For example, the distance probed in the LHC experiments is of the order of $10^{-20}$m, while $l_P \approx 10^{-34}\textrm{m}$, so that we can have 
\be 10^{-20}\textrm{m} > |L_\e| \gg 10^{-34}\textrm{m} \, . \ee

One can also add the cosmological constant (CC) term to the Regge action, so that
\be S_R(L) \to S_R(L) + \L_c V_4 (L) \, . \ee
The addition of the CC term does not affect the convergence properties of the path integral, so that we still have the same bound for the parameter $p$ as in the zero CC case. However, the condition for the semi-classical expansion of the effective action for $\L_c = 0$, given by $|L_\e| \gg l_P$ and $L_0 \gg l_P$, is replaced by the condition
\be |L_\e| \gg l_P \,,\quad  L_0 \gg \sqrt{l_P L_c} \,,\ee
where $|\L_c| = 1/L_c^2$ \cite{MV}. 

It is important to show that a QG theory has the spectrum of the cosmological constant such that it includes the observed value, see \cite{Pcc}. In the PFQG case, it can be shown that the observed value of the CC belongs to the allowed interval of values, provided that the path integral for gravity plus matter is finite, see \cite{MV, MVb}. Since the path integral is finite for $p>$ 52,5 then the proof given in \cite{MVb} is now complete. Note that in the string theory case, this is a more difficult problem, since the string CC spectrum is discrete and almost all values are negative \cite{Pcc}.

\section{Conclusions}

A PFQG theory defined by the path integral (\ref{gmpi}) with the PI measure (\ref{pim}) and the bound (\ref{prm}) is the first example of a mathematically complete theory of quantum gravity which includes the SM matter. By mathematically complete QG theory we mean that all the transition amplitudes are well-defined and finite and that it has the classical limit which is given by GR coupled to the SM. 


Whether a PFQG theory is realized in Nature or not, remains to be seen, because this depends on the existence of phenomena that are unique to PFQG. Also, if there are phenomena that cannot be explained by PFQG, this would mean its rejection or a modification. So far, all the known phenomena which can be atributed to quantization of gravity, like the value of the CC, Big Bang, inflation and evaporation of black holes can be in principle explained by PFQG. A specific PFQG effect would be the appearence of deviations from the QFT scattering amplitudes at high energies due to the non-smooth nature of the spacetime.

The physics of the PFQG theory in the $\S\times I$ part of the spacetime is described by the dynamics of the effective action. An EA trajectory can describe a vacuum expectation value trajectory, provided that the inital data are chosen appropriatelly while the analog of the QFT vacuum state is the HH state. 

The perturbative effective action can be obtained via the long edge-length ($|L_\e| \gg l_P$) approximation.  For a large number of the edge lengths, which are microscopically small, one obtains the smooth-manifold approximation and the corresponding QFT has the cutoff $K$ which is determined by the average edge length. When $\hbar K \ll E_P$, the perturbative QFT effective action is a good approximation for the perturbative PFQG effective action.

It is reasonable to expect that the new physics will be described by the non-perturbative effective action, i.e. when $|L_\e| \approx l_P$. In this case
one can use the short edge-length expansion of the effective action
\be \G(L,\F) \approx \sum_{k_1 \ge\, 0,\,  k_2 \ge\, 0,\,  \cdots\,, \,k_N \ge\, 0} (l_P)^{-(k_1 + \cdots +k_N)} \,\g_{k_1 k_2 \cdots k_N} (\F) L_1^{k_1} L_2^{k_2} \cdots  L_N^{k_N}\,, \ee
where the coefficents $\g$ can be obtained by substituting the above expansion into the EA equation.

One can also try to obtain $\g(\F)$ functions by calculating the generating function $Z(J,j)$, i.e. the path integral with the edge-length currents $J$ and the matter currents $j$.  The effective action can be then obtained by performing the Legandre transform
\be \G( L, \F) = W(J,j) - J L - j \F \,, \quad  L =  {\pa W\over \pa J}\,, \quad \F =  {\pa W\over \pa j} \,,\label{npea}\ee
where $W = -i\hbar \log Z$. This approach requires solving the last two equations so that $J$ and $j$ are expressed as functions of $L$ and $\F$. This calculation will simplify in the limit $L \to 0$ and  $\F\to 0$.

Since the PFQG effective action is very well approximated by the perturbative QFT effective action for a smooth spacetime when the edge lengths are microscopically small in $T(\S\times I)$, but larger than the Planck length, we then have
\bea \G &\approx& \left(\frac{1}{G_N} + a_1\hbar K^2 \right) S_{EH} + \left(\L_c + \hbar b_1 K^4 \ln \left({K\over k_0}\right)  \right) \int_{\S\times I} d^4 x \sqrt{|g|} + S_m(m_1,g_{1}) \cr
&+&\,\hbar\ln\left({K\over k_0}\right) \int_{\S\times I}\sqrt{|g|}\left( c_1 R^2 + d_1 R_{\m\n}R^{\m\n} + e_1 R_{\m\n\r\s}R^{\m\n\r\s} + f_1 \nabla^2 R \right) d^4 x \,, \label{smolea}\eea
where $S_m(m_1,g_{1})$ is the matter action for the one-loop corrected matter coupling constants and masses. We have also neglected the $O(\hbar^2)$ EA terms and $O(\nabla^6)$ terms in the one-loop EA contribution \cite{MV}. The $O(\nabla^4)$ terms, which include the quadratic in curvature terms, are the terms that are responsible for the Starobinsky inflation \cite{Strb}. Hence the Starobinsky inflation can be naturally realized in the PFQG theory.

From the expression (\ref{smolea})  we can see that $G_N$ and $\L$  will run with the cutoff $K$ as
\be  \frac{1}{G_N^*} =  \frac{1}{G_N} +  a_1\hbar K^2 + \cdots \,,\quad  \L = \L_c + \hbar b_1 K^4 \ln \left({K\over k_0}\right)+ \cdots  \,.\ee
These corrections will blow up as $K\to\infty$, but this is not a problem, since these expansions are valid only in the the perturbative QFT approximation, for which $\hbar K \ll E_P = O(10^{19})$ GeV. 

The exact values of $G_N^*$ and $\L$ can be obtained by using the PFQG effective action $\G(L,\F)$,
which is determined by the EA equation (\ref{gmea}). Since the EA equation is a consequence of the defining relations (\ref{npea}), then the EA equation is well-defined only when the path integral is finite\footnote{Note that the perturbative solution of the EA equation does not depend on the convergence properties of the path integral. Hence the perturbative solution exists even when the path integral is divergent, see \cite{M,MV}.}. Consequently,  for $p>52,5$ one can construct a nonperturbative effective action by using the equation (\ref{npea}).  

In particular, the existence of the non-perturbative effective action implies $G_N^* = G_N$, i.e. $G_N$ does not have quantum corrections,  while 
\be\L = \L_c + {l_P^2 \over 2 L_0^4} + \varepsilon (m_0, g_0, l_P^2) \,,\ee
where $m_0$ and $g_0$ are the low-energy values of the SM masses and couplings, see \cite{MV, MVb}. Note that $\L_c$ is a free parameter, and the quantum corrections are given by a contrubution from the PI measure plus a contribution from the matter vacuum energy density, which is finite and it is determined by the non-perturbative effective action.

Our last remark is that the PFQG theory generalizes Quantum Mechanics and QFT in the following way:

i) the initial state of the Universe is given by the HH state, which is determined by the ``vacuum'' 4-manifold $M$ of Fig. 3 and a chosen triangulation,

ii) the WFU time evolution is given by the path integral for $M \sqcup \left(\S\times [0,t] \right)$ and the corresponding triangulation,

iii) the corresponding effective action is determined by the manifold of Fig. 4, so that a solution of the EA equations of motion corresponds to the expectation values of the Heisenberg picture operators for the metric and the matter fields in the initial state.

\end{document}